\documentclass[conference]{IEEEtran}

  	\usepackage[pdftex]{graphicx}
  	\graphicspath{{../pdf/}{../jpeg/}}
	\DeclareGraphicsExtensions{.pdf,.jpeg,.png}

	\usepackage[cmex10]{amsmath}
	\usepackage{mathabx}
    \usepackage{hyperref}
    \usepackage{algorithm}
    \usepackage{algpseudocode}
	\usepackage{array}
	\usepackage{mdwmath}
	\usepackage{mdwtab}
	\usepackage{eqparbox}
	\usepackage{url}
    \usepackage{comment}
        \usepackage{subfigure}
        \usepackage{romannum}
        \usepackage{comment}

\begin{document}





\title{Using Early Exits for Fast Inference in \\Automatic Modulation Classification}


\author{\IEEEauthorblockN{Elsayed Mohammed, Omar Mashaal, and Hatem Abou-Zeid}

\IEEEauthorblockA{{Department of Electrical and Software Engineering}, 
{University of Calgary},
Calgary, Canada\\ \{elsayed.mohammed, omar.mashaal1, hatem.abouzeid\}@ucalgary.ca }

\thanks{This research was supported by the Natural Sciences and Engineering Research Council of Canada (NSERC) under Grant RGPIN-2021-04050.}

}

\maketitle

\begin{abstract}
Automatic modulation classification (AMC) plays a critical role in wireless communications by autonomously classifying signals transmitted over the radio spectrum.
Deep learning (DL) techniques are increasingly being used for AMC due to their ability to extract complex wireless signal features.
However, DL models are computationally intensive and 
incur high inference latencies.
This paper proposes the application of early exiting (EE) techniques for DL models used for AMC to accelerate inference.  
We present and analyze four early exiting architectures and a customized multi-branch training algorithm for this problem. 
Through extensive experimentation, we show that signals with moderate to high signal-to-noise ratios (SNRs) are easier to classify, do not require deep architectures, and can therefore leverage the proposed EE architectures. 
Our experimental results demonstrate that EE techniques can significantly reduce the inference speed of deep neural networks without sacrificing classification accuracy. 
We also thoroughly study the trade-off between classification accuracy and inference time when using these architectures.
To the best of our knowledge, this work represents the first attempt to apply early exiting methods to AMC, providing a foundation for future research in this area.




\end{abstract}

\IEEEoverridecommandlockouts

\IEEEpeerreviewmaketitle


\section{Introduction}

Automatic modulation classification (AMC) plays a critical role in wireless communications by autonomously classifying signals transmitted over the radio spectrum. It facilitates a range of applications such as: interference detection and monitoring, dynamic spectrum access, and enhanced target recognition in radar systems. AMC is also of great interest to regulatory and defense organizations as it enables the detection of illegal and malicious spectrum usage. AMC is therefore of fundamental importance as  beyond 5G and 6G applications focus on connectivity of critical infrastructure. 



Deep learning (DL) models have enabled accurate AMC and several DL architectures have been developed \cite{firstCNN},\cite{overTheAir}. The success of DL is due to its ability to automatically extract complex features from the wireless signals. 
However, while DL has demonstrated high accuracy, the models used are typically large and require significant computational resources. This results in high energy consumption and a large inference time. This 
is a significant limitation when the wireless system has to react to the detected modulation in a timely fashion. 

However, complex DL models with a large inference time and energy consumption may be unnecessary for signals received at high signal-to-noise (SNR) levels.  
Our intuition is that signals with a high SNR should be easier to classify than signals with a low SNR. In such cases, it may be sufficient to utilize shallower DL architectures that require significantly less computational resources. However, those shallow architectures may not succeed when tasked to classify signals with low SNRs. 
This motivates the question of whether deep learning architectures with \emph{multiple} branches can be used to address this dilemma? The goal would be for signals with high SNRs to \emph{automatically} branch out after traversing a few layers of the network $-$ while those with low SNRs continue throughout the full deep neural network to determine the AMC result. 

\begin{figure}[t!] 
\centering
\includegraphics[width=1.6in]{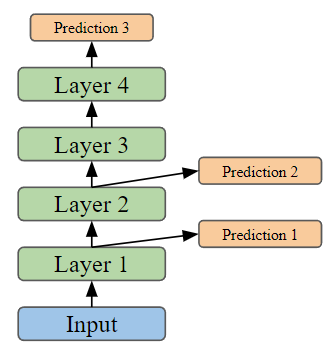}
\caption{Demonstration of Early Exiting Inference.}
\label{eeFig}
\end{figure}

Indeed, an approach known as early exiting (EE) has been recently introduced in the deep learning literature to reduce the latency of large DL models \cite{branchynet}. 
EE reduces the latency for samples that can be easily classified by an intermediate classifier without going through all the inference path as shown in Figure \ref{eeFig}. All samples will first execute the first branch or classifier labeled as Prediction 1 in Figure \ref{eeFig}. The decision to accept the classification made at this intermediate classifier is based on an entropy measure that quantifies the model's confidence of the output. If the entropy is above a threshold, the output is not accepted and the remaining branches are executed in succession.  


Our work therefore aims to propose and investigate the utility of EE for AMC. In this paper, we hypothesize that EE may achieve a comparable classification accuracy to traditional DL models while providing a significant reduction in inference latency, particularly at higher SNRs. In more detail, the main contributions and insights of this paper are:
\begin{itemize}
    
    \item We propose the adoption of early exiting-based CNN architectures to enable fast inference in AMC. To the best of our knowledge, this work represents the first attempt to apply early exiting methods to AMC, providing a foundation for future research in this area. 
    \item Our work presents four early exiting architectures and a customized multi-branch training algorithm for this problem. Through extensive experimentation, we demonstrate the effectiveness of applying early exiting to achieve high accuracy and low inference time in AMC for signals at different SNR levels.
    We show that this requires careful design of the EE training process, location of the EE in the architecture, and the values of the entropy confidence thresholds. A thorough evaluation study and discussion is presented to illustrate the latency-accuracy trade-offs, and provide guidelines to the design of EE for AMC.
    \item We have conducted a thorough investigation of the performance of EE with different SNRs. Our results validate the hypothesis that signals with higher SNRs: 1) have a higher likelihood of terminating at the early exits, and 2) are indeed classified with a high accuracy at the early exits and do not require deep architectures. 
\end{itemize}

The results of this paper highlight the significant potential that EE deep learning architectures can have to enable low-latency inference in AI-powered wireless communications. All the code to reproduce our results and conduct further research in this area 
is available at \url{https://t.ly/2vSID}.

The rest of this paper is organized as follows. Section \Romannum{2} provides a summary of related work. In section \Romannum{3}, we describe our system model and the dataset used. Following that, the methodology of the proposed EE architectures is presented in Section \Romannum{4}. Section \Romannum{5} then discusses our results and findings, and we conclude the paper in Section \Romannum{6}.


\section{Related Work}

In this section, we briefly discuss key works on AMC and EE, independently. 
To our knowledge EE has not been investigated in the context of AMC.

\textbf{Automatic Modulation Classification.}
AMC has been thoroughly investigated using traditional machine learning \cite{ACMbyML}, signal processing techniques, and feature-extraction methods \cite{cyclicAMC}. However, some of the challenges faced are the requirements of high sampling rates and sensitivities to the selected features.  
More recently, deep learning has emerged as a robust alternative and impressive results have been reported using several CNN-based architectures \cite{firstCNN}, \cite{2DCNN}, \cite{overTheAir}. The work in \cite{Xnet} proposed a convolutional auto-encoder design to recognize the modulation type with a focus on improving robustness to noise. More complex composite DL architectures have also been proposed in \cite{withPCA} to improve the ability to accurately classify each modulation type.  

\textbf{Early Exiting in Deep Learning.}
Early exiting for deep neural networks was proposed for the first time as a solution for the model overthinking problem as well as to increase the inference speed \cite{branchynet}. In the proposed branchyNet architecture \cite{branchynet}, multiple side branches are augmented from the main architecture. These branches are used to classify a number of input samples without going through the whole network. 

Several approaches have been introduced in literature to train the EE architectures. One approach is to train the entire architecture as a single entity by  formulating a unified optimization problem that takes into account all intermediate exits \cite{EEreview}. This can be accomplished through a variety of methods, such as using per-classifier losses \cite{train1} or by combining the intermediate outputs before applying a single loss function \cite{train2}. Other approaches have also been proposed where the common network and exit branches are trained separately.


Early Exiting has been successfully applied in different tasks to accelerate inference without significant degradation in performance. For example, it has been used in natural language processing (NLP) tasks to improve the efficiency of models that process large amounts of text. DeeBERT \cite{DeeBERT} is a recent study that successfully demonstrated the efficiency of applying early exiting with large-scale language models such as BERT. 


\section{System Model}

\subsection{Signal Model and Problem Statement}
In wireless communications, the received signal \(r(t)\) is the result of channel effects on the transmitted modulated signal \(s(t)\) including distortions of signal’s amplitude and spectrum. Thus, the received signal in a wireless system is described by the following formula:
\[r(t) = s(t) * h(t) + n(t)\]
where \(h(t)\) is the wireless channel frequency response and \(n(t)\) is the additive noise.


Given the received signal at the receiver, the task of AMC is to identify the modulating type of the transmitted one. The problem then follows the definition of the maximum-a-posterior (MAP) criterion. 
\[m_{i} = \arg \max_{m_{i} \in M}{P(m_{i}|r)}\]
where \(M = \{m_i\}_{i=1}^{N}\) and \(N\) is the number of modulation techniques \cite{lightAMC}.

\subsection{Dataset}

We use the recently released RML22 dataset \cite{RML22} in our study. 
RML22 is an open-source dataset created based on the benchmark dataset RADIOML.2016.10A (RML16)
and addresseses some of its shortcomings. Like RML16, the RML22 dataset is used as a synthetic dataset for different wireless problems including but not limited to AMC.  

The dataset contains 420000 received baseband signals in 10 different modulation techniques: {BPSK, QPSK, 8PSK, 16QAM, 64QAM, PAM4, CPFSK, GFSK, WBFM, AM-DSB}. The dataset is evenly distributed among the 10 modulation techniques and 21 SNR levels from –20 dB to 20 dB in steps of 2 dB. Therefore, the dataset has 2000 data points (samples) per modulation type per SNR level. Each sample is in IQ format with a size of (2, 128). We use 10\% of the dataset to test our AMC models, and 90\% for model training and validation (10\% of the 90\% is used for validation).
This comes to 340200 samples for training, 37800 for validation, and 42000 are allocated for testing.

\section{Early Exiting DL Architectures for AMC}

In this section, we provide a detailed description of the proposed early exiting architectures, and their associated training and inference algorithms. A description of the backbone deep learning model used is also first presented.

\subsection{Baseline Model}

\begin{figure}[b!] 
\centering
\includegraphics[width=0.8in, angle =-90]{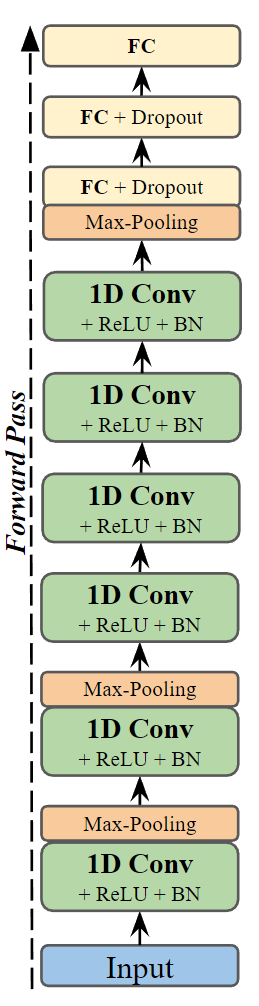}
\caption{Baseline Backbone Model Architecture for AMC.}
\label{BLModelFig}
\end{figure}

To evaluate the effectiveness of early exiting in AMC, we first implemented a high-performing benchmark baseline network, based on the 1D-CNN architecture described in \cite{RML22} and shown in Figure \ref{BLModelFig}. 
The baseline backbone architecture leverages the benefits of CNNs, such as parameter sharing, translation invariance, and hierarchical representation \cite{DLref}. Based on the network designed in \cite{RML22} for the RML22 modulation classification dataset, our baseline model utilizes 1D convolutional layers for core feature extraction. The architecture, shown in Figure \ref{BLModelFig}, consists of six consecutive 1D convolutional layers. Each convolutional layer in the baseline architecture extracts distinct features for accurate modulation classification. To enhance the network's ability to capture complex representations, a Rectified Linear Unit (ReLU) activation function is applied after every convolutional layer. Additionally, max-pooling and batch normalization layers are inserted after some of these layers, facilitating dimension reduction and improving model convergence. The final stage consists of a max-pooling layer, 3 fully connected (FC) layers, and a SoftMax activation, producing an output of size (1, 10) representing the number of modulation types. The SoftMax layer ensures that each neuron's output represents the probability of the input sample belonging to each class. To prevent overfitting, an implicit dropout layer with a dropout rate of 0.3 is included after each FC layer, except for the last layer.

\subsection{Proposed Early Exiting Architectures}

The proposed early exiting architecture aims to achieve fast inference without compromising classification accuracy across all signal to noise ratios. 
The goal is to achieve a significant reduction in computations by exiting the backbone network early $-$ but to do so only when the received signal can be classified accurately with an early exit. To accomplish this, there are two main design aspects to be considered: 1) where the early exit branches off from the backbone network, and 2) the threshold value of the entropy confidence level that will terminate the inference path.



\begin{figure*}[t!] 
  \centering
  \subfigure[$V0$]{\includegraphics[width=0.21\textwidth]{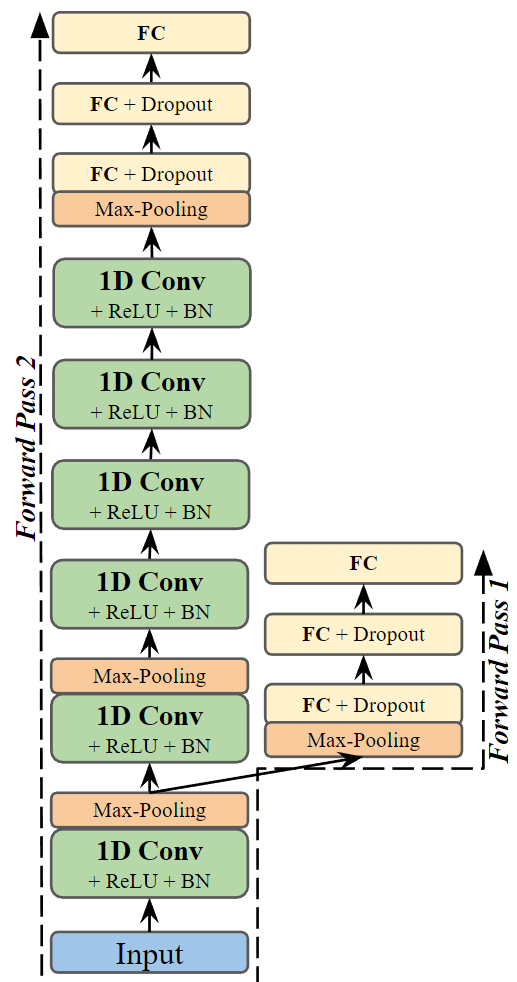}\label{V0Fig}}
  \subfigure[$V1$]{\includegraphics[width=0.21\textwidth]{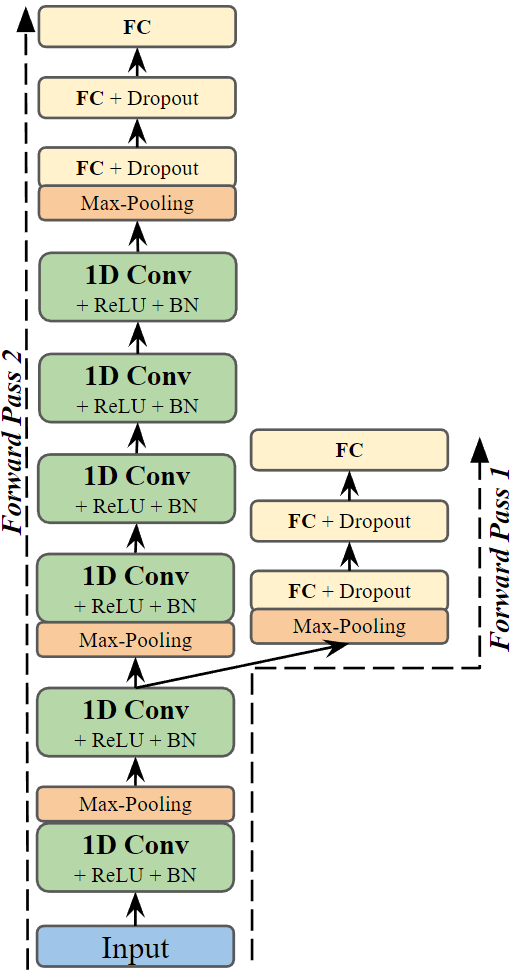}\label{V1Fig}}
  \subfigure[$V2$]{\includegraphics[width=0.21\textwidth]{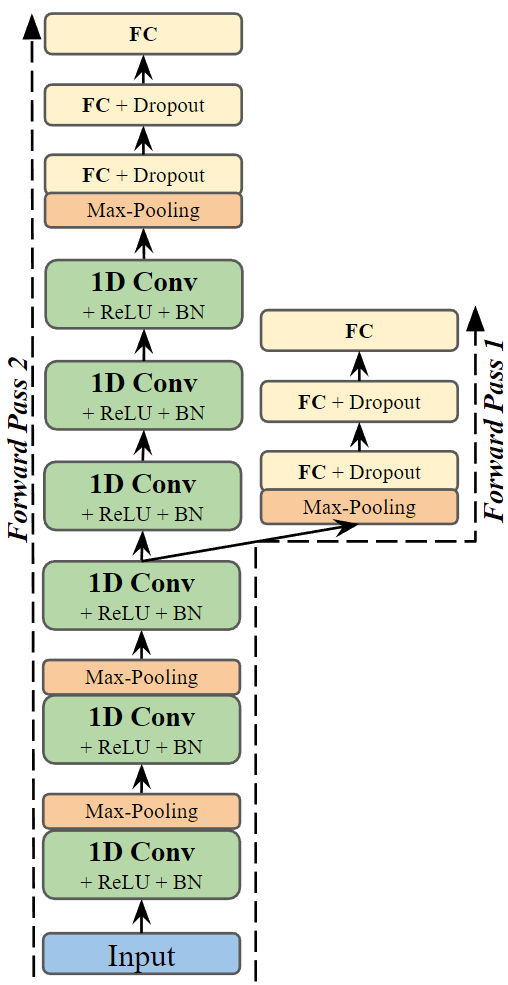}\label{V2Fig}}
  \subfigure[$V3$]{\includegraphics[width=0.21\textwidth]{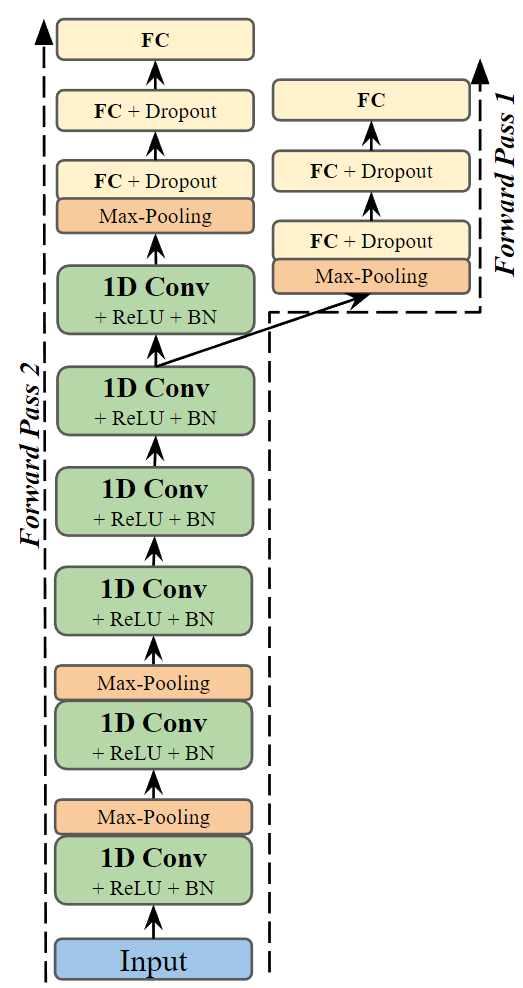}\label{V3Fig}}
  \caption{Proposed Early Exiting Model Architectures for AMC Classification.}
  \label{modelsFig}
\end{figure*}

We therefore implement early exits at four different locations of the backbone network to investigate the impacts and trade-offs that each architecture provides.
These EE architectures are illustrated in Figure \ref{modelsFig} and are denoted with the naming convention (V0 to V3) based on their proximity to the input, with V0 being the closest and V3 being the farthest. 
All EE architectures maintain the same number and types of layers for the backbone architecture $-$ and the exit branches have identical intermediate classifier architectures. This allows for a comparative analysis of the impact of early exiting location on the overall performance of the network.



\begin{algorithm}[t!]
\caption{Proposed Early Exiting Training}\label{EE_Training_Algorithm}
\textbf{Inputs:} {Training data $X$, target labels $y$, loss function $\mathcal{L}$ optimizer $\mathcal{O}$, number of epochs $E$} 

\textbf{Outputs:} {Trained network parameters $\theta_1, \theta_2$}

// $\theta_1$ : \text{$Forward\ Pass\ 1$ layers}

// $\theta_2$ : \text{$Forward\ Pass\ 2$ layers except for the common layers}

\begin{algorithmic}
\Procedure{Training}{$X$, $y$}
\State Initialize network parameters $\theta_1$, $\theta_2$\;
\For{$epoch$=1 $\to$ $E$}
\For{$batch$ in $X$}
\State \parbox[t]{\dimexpr\textwidth-\leftmargin-\labelsep-\labelwidth} {$Forward\ Pass\ 1$: compute network output $\hat{y_1}$ \strut} 
\State \parbox[t]{\dimexpr\textwidth-\leftmargin-\labelsep-\labelwidth} {$Forward\ Pass\ 2$: compute network output $\hat{y_2}$ \strut} 
\State Compute $Loss\ 1$: ${loss_1} \gets $$\mathcal{L}(\hat{y_1},y)$\;
\State Compute gradients 1: $ \nabla_1 \gets \nabla_{\theta_1} \mathcal{L}(loss_1)$\;
\State Update parameters $\theta_1 \gets \mathcal{O}(\theta_1, \nabla_1)$\;
\State Compute $Loss\ 2$: ${loss_2} \gets $$\mathcal{L}(\hat{y_2},y)$\;
\State Compute gradients 2: $ \nabla_2 \gets \nabla_{\theta_2} \mathcal{L}(loss_2)$\;
\State Update parameters $\theta_2 \gets \mathcal{O}(\theta_2, \nabla_2)$\;
\EndFor
\EndFor
\State  \textbf{return} $\theta_1$, $\theta_2$
\EndProcedure

\end{algorithmic}

\end{algorithm}

\subsection{Early Exit Training Engine}

Early-exiting architectures have two or more branches that share a common set of network layers. As shown in Figure \ref{modelsFig}, the further away the exit or branch is from the input, the larger the number of shared network layers. Neural network training in EE architectures is therefore different from traditional networks since it has to incorporate a forward pass per branch and a resulting loss per branch.  

In our proposed architectures, the training process involves two forward passes, as illustrated in Figure \ref{modelsFig}.
The common network layers and the early exiting branch is referred to as \textit{Forward Pass 1} 
and full backbone branch is referred to as \textit{Forward Pass 2}.
Consequently, two losses are calculated: \textit{Loss 1} 
measures the cross-entropy between the output of \textit{Forward Pass 1} and the ground truth labels, while  \textit{Loss 2}  quantifies the cross-entropy between the output of \textit{Forward Pass 2} and the ground truth labels. To update the model's parameters, \textit{Loss 1} is used for the backpropagation computations of gradients of the common layers and the early exit layers. On the other hand, \textit{Loss 2} is only used for the backpropagation computations of gradients for the main backbone network layers, excluding the common layers. In other words, \textit{Loss 2} only affects the layers starting from the branching location up to the end of the backbone network. This training procedure for the early exiting architectures is summarized in Algorithm \ref{EE_Training_Algorithm}.

\subsection{Early Exit Inference Engine}


The inference engine is responsible to execute the forward passes of the multi-branch architecture after training has converged and to determine whether an input sample can terminate at an early exit. The primary motive to use EE architectures in classification is to reduce the computational burden and inference time for input samples that can be correctly classified with an early exit or short branch. 

To classify an input sample, it first traverses the initial common layers before reaching the first branch. The intermediate output of these layers, denoted as Q, is then saved and forwarded to the exit branch where the fully connected layers are executed. The EE resulting output, referred to as $z_i$ (where $i$ is the exit index), undergoes an entropy evaluation to determine the confidence level of this classification as follows:
\[E({z_i}) = \sum_{i=0}^{9}{-p_i \log_{10}{p_i}}\]
This confidence level is then compared against a threshold value, $T$, which determines whether the classification $z_i$ will be accepted or rejected. If $E(z_i)$ is less than or equal to $T$, the temporary classification $z_i$ becomes the model's inference output. However, if $E(z_i)$ exceeds $T$, the inference engine will redirect the input to the main backbone branch to be classified by \textit{Forward Pass 2}. 
This inference procedure for our early exiting architectures is summarized in Algorithm. \ref{EE_Inference_Algorithm2}.

\begin{algorithm}[t!]
\caption{Proposed Inference Algorithm}\label{EE_Inference_Algorithm2}

\textbf{Inputs:} {Data $X$, Confidence threshold $T$}

\textbf{Outputs:} {Predicted labels $\hat{y}$}

\begin{algorithmic}
\Procedure{Inference}{$X$, $T$}
\For{$sample$ in $X$}
    \State \parbox[t]{\dimexpr\textwidth-\leftmargin-\labelsep-\labelwidth} {$Forward\ Pass\ 1$: compute network output $\hat{z_1}$\strut}
    \State { Common layer's output $Q$ is saved}
    \If{$entropy(z_1) < T$}
    \State $\textbf{append}(\hat{y}, z_1)$\; 
    \Else{}
    \State{$Forward\ Pass\ 2$ begins with $Q$}
    \State \parbox[t]{\dimexpr\textwidth-\leftmargin-\labelsep-\labelwidth} {$Forward\ Pass\ 2$: compute network output $\hat{z_2}$ \strut}
    \State $\textbf{append}(\hat{y}, z_2)$\; 
    \EndIf

\EndFor
\State  \textbf{return} $\hat{y}$
\EndProcedure
\end{algorithmic}
\end{algorithm}

\begin{figure*}[hbtp!]
  \centering
  \subfigure[Accuracy]{\includegraphics[width=0.32\textwidth]{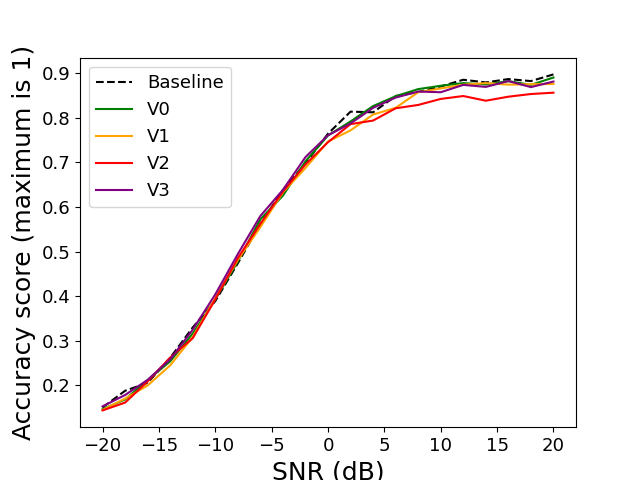}\label{accuracyAt0.35}}
  \subfigure[Early Inference Percentage]{\includegraphics[width=0.32\textwidth]{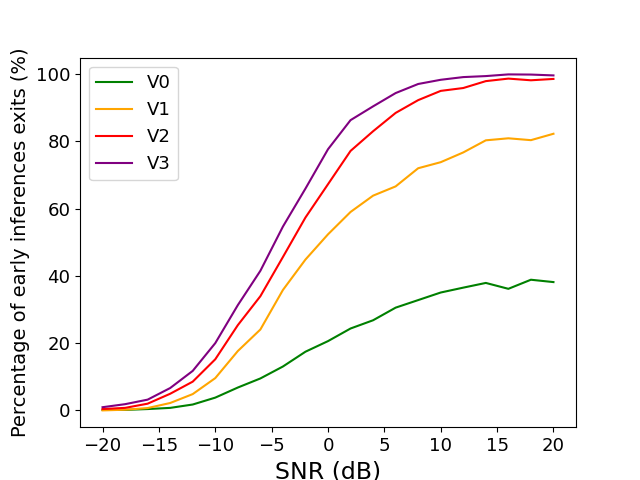}\label{infPercentage0.35}}
  \subfigure[Inference Time]{\includegraphics[width=0.32\textwidth]{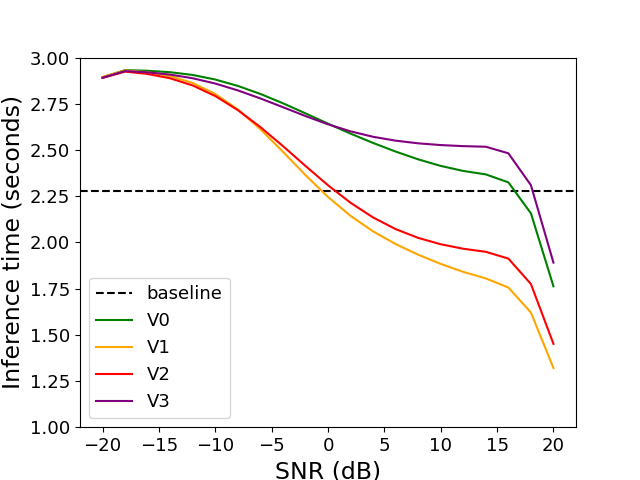}\label{infTime0.35}}
   \caption{Percentage of early inference exits and the corresponding average inference time at different SNRs. The entropy threshold $T=0.35$.}
  \label{inferenceFigure0.35}
\end{figure*}



\section{Results and Discussion}
All the four EE architectures were trained and evaluated using the EE algorithms presented in Section IV. 
We now present and discuss the results obtained from the test set to evaluate the effectiveness of EE in reducing computational complexity and achieving high classification accuracy  on unseen data.
The experimental setup utilized a 12th Gen Intel(R) Core(TM) i9-12900K/3.20 GHz processor with 64.0 GB RAM and an NVIDIA GeForce RTX 3080 Ti graphics card. PyTorch is the main framework used for code development. 

\subsection{Early Exit Architectures - Effect of Exit Location}
We first thoroughly study the effect of the \emph{location} of the early exit (i.e., the architecture choice) on the classification accuracy, the frequency of exiting early, and the resulting inference time. These results are for a confidence threshold ($T=0.35$) that was shown to provide a good trade-off between accuracy and inference time. In the next subsection, we will discuss the effects of this threshold.

\textbf{Classification Accuracy.} 
Figure \ref{accuracyAt0.35} shows the accuracy of the baseline backbone model and the four proposed EE architectures of Figure \ref{modelsFig}. The results follow the expected AMC trends where accuracy increases with increasing SNRs. All the models perform almost identically for low SNRs. Results vary amongst the models for higher SNR values ($>0$  dB) but we observe that $V0$, $V1$, and $V3$ perform very close to the baseline backbone model. To provide more insights on these results, we discuss the percentage of early inference exits next. 

\textbf{Frequency of Early Exits.} 
Figure \ref{infPercentage0.35} shows the percentage of samples that were classified using the early exit branches for the different architectures. We observe two trends that validate our hypothesis that motivated the study of EE for AMC. The first that is indeed higher SNRs result in a higher percentage of samples that are successfully classified with the EE branches. It is therefore unnecessary to design very deep NN architectures when the SNR exceeds $0$ dB. This also indicates that the entropy measure in the inference engine captures the degree of classification certainty in AMC $-$ and that this certainty increases with increasing SNRs as one would expect. The second observation from Figure \ref{infPercentage0.35} is that when early exits happen very close to the input as in model $V0$, the entropy measure is too high and the percentage of successful classifications at this exit is low. This is expected since a very shallow architecture may not be sufficient. It is however worth noting that while $V0$ and $V1$ differ by a single 1D-Conv layer, this addition is enough to result in much larger \%-age of exits. 


\begin{figure}[t!]
  \centering
  \subfigure{\includegraphics[width=0.42\textwidth]{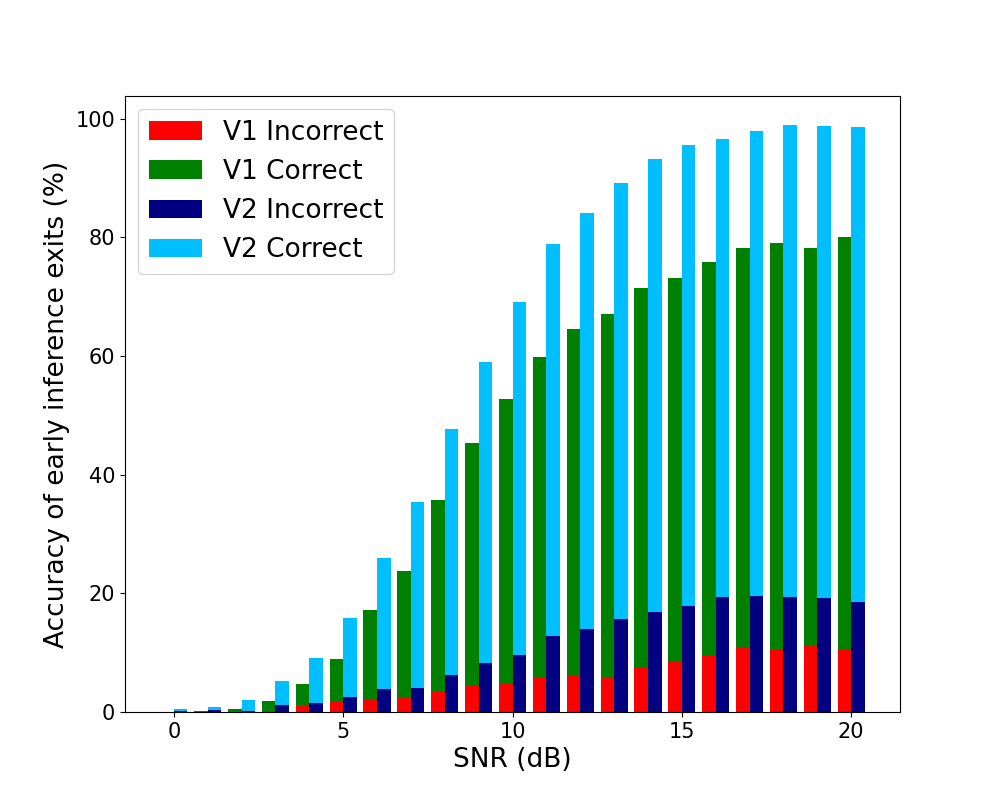} }
  \caption{Comparison of accuracy for the $V1$ and $V2$ EE at $T=0.35$.}
  \label{stackedBar}
\end{figure}

A rather interesting result of Figure \ref{infPercentage0.35} is that $V2$ has a very high \%-age of early exits, but Figure \ref{accuracyAt0.35} indicates the accuracy of $V2$ has taken a small hit compared to the rest of the architectures. To investigate if this was due to the mis-classifications of the $V2$ early exit, we plot the EE accuracies for $V1$ and $V2$ in Figure \ref{stackedBar}. The figure shows that indeed exits at $V2$ have a considerably higher percentage of incorrectly classified modulations, and this has resulted in the accuracy drop observed in Figure \ref{accuracyAt0.35}. As such, careful considerations should be made to the location of the EE branch since it can have an impact on the resulting performance. 


\textbf{Inference Time.}
In order to show the benefits of the proposed EE strategies, we plot the resulting inference times of the different architectures in Figure \ref{infTime0.35}. Architectures $V0$ and $V3$ do not provide inference time gains except at very high SNRs and this is expected due to different reasons. $V0$ is a very shallow architecture and most of the samples are not classified after traversing the EE branch due to a high entropy. This results in an additional  delay incurred by using this branch. $V3$ is a very long architecture and despite that, at low to medium SNR values, the entropy values are not low enough for a successful exit. The entropy calculation itself introduces additional inference delays for even cases where EE is successful. Both $V0$ and $V3$ were designed as extreme branching locations and are provided mainly to validate and gauge the relative behavior of EE architectures in AMC. 

Figure \ref{infTime0.35} shows that $V1$ and $V2$ have considerably lower inference times than the baseline architecture for positive SNR values. This demonstrates the value of EE-based designs for AMC. Further optimizations are also possible. For example, the entropy calculation does not need to happen at every signal sample once a signals SNR is deemed suitable for the early exit. To reduce the inference time of the SNR signals, they may be explicitly directed toward the back-bone branch and EE may be disabled based on their SNR value. Adaptive entropy thresholds can also be introduced for low SNR signals to allow them to exit early despite a high entropy. This is because they are unlikely to be classified accurately even by the long branch. 
Inference time can also be further reduced by calibrating the entropy threshold as discussed next. 




\begin{figure}[t!]
  \centering
  \subfigure[$T=0.6$]{\includegraphics[width=0.23\textwidth]{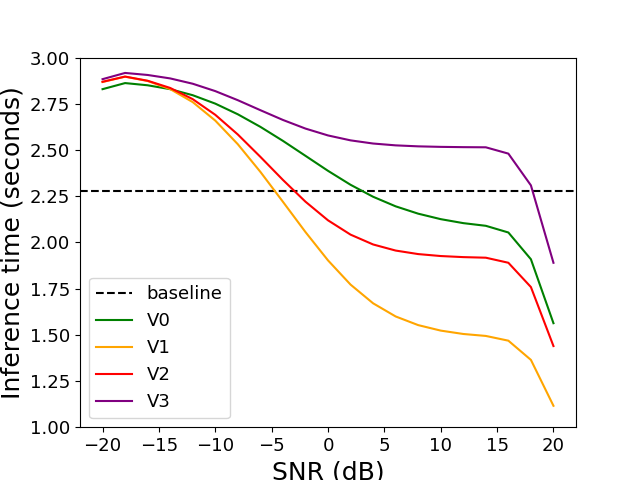} \label{timeAt0.6}}
   \subfigure[$T=0.05$]{\includegraphics[width=0.23\textwidth]{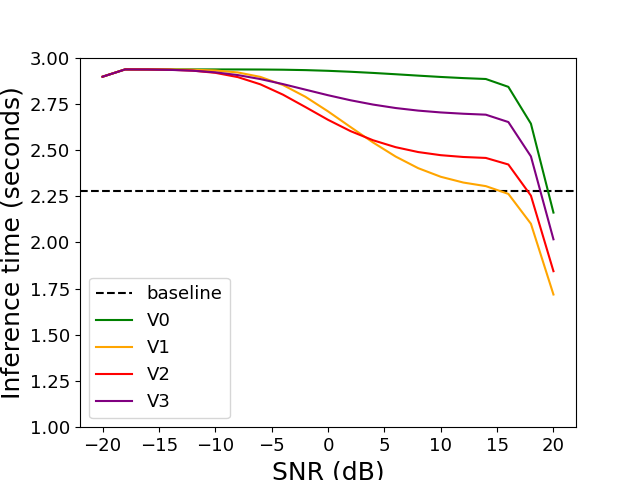} \label{timeAt0.05}}
  \subfigure[$T=0.6$]{\includegraphics[width=0.23\textwidth]{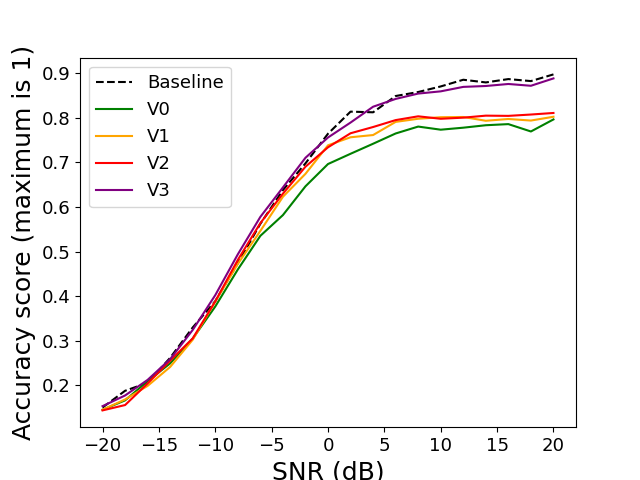} \label{accuracyAt0.6}}
   \subfigure[$T=0.05$]{\includegraphics[width=0.23\textwidth]{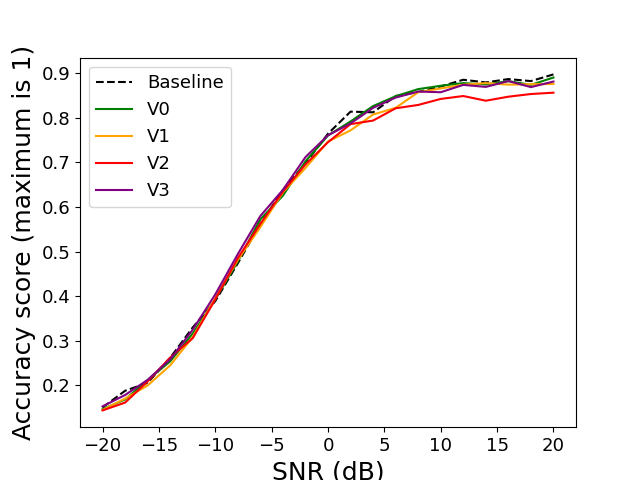} \label{accuracyAt0.05}}
  \caption{Inference time and accuracy results for the proposed early exit models at different SNR levels with  $T=0.6$ and $T=0.05$.}
  \label{accuracyAt0.4and0.05}
\end{figure}
\subsection{Entropy Threshold: Accuracy-Latency Trade-offs}
Using a higher entropy threshold (lower confidence level) will result in more frequent early exits for all architectures. This is illustrated in Figure \ref{timeAt0.6} that shows the considerable decrease in inference time, particularly for $V1$ amongst all the architectures due to its shallow but effective architecture. This inference time reduction does however compromise the accuracy as shown in Figure \ref{accuracyAt0.6}. It is worth noting that $V3$ does not undergo the same accuracy reduction since it has a far EE and most inputs are classified correctly. This comes at the cost of a much larger inference time. 

Results for reducing the entropy threshold are also provided in Figure \ref{timeAt0.05} and Figure \ref{accuracyAt0.05}. The inference time is now increased since most inputs will not successfully exit the branches with this threshold and will be redirected to traverse the full backbone path as well. These results demonstrate the importance of calibrating the entropy threshold to achieve the desired accuracy-inference time trade-offs. 

\noindent\textbf{Summary.} 
These results demonstrate that the proposed EE architectures can achieve high AMC accuracies while providing significantly lower inference times. By studying these trade-offs between accuracy and inference time, we can conclude that using $V1$ is the most suitable amongst the studied architectures. Further optimizations are also possible for additional reductions to the inference time.

\section{Conclusion}
This paper explored the application of early exiting techniques for automatic modulation classification. Our work presented four early exiting architectures and a customized multi-branch training algorithm for this problem. Through extensive experimentation, we demonstrated the effectiveness of applying early exiting to achieve high accuracy and low inference time to classify modulations for signals at different SNR levels. We discussed the inherent trade-offs between classification accuracy and inference latency when using these architectures. To our knowledge, this work represents the first exploration of applying early exiting methods to automatic modulation classification, thus paving the way for further investigation in this domain. Future research could focus on developing adaptive entropy thresholds and customized training and inference algorithms to improve the performance and efficiency of the proposed early exiting techniques.






\bibliographystyle{ieeetr}
\bibliography{IEEEabrv,main}

\begin{thebibliography}{10}

\bibitem{firstCNN}
T.~J. O’Shea, J.~Corgan, and T.~C. Clancy, ``Convolutional radio modulation
  recognition networks,'' in {\em Engineering Applications of Neural Networks},
  pp.~213--226, Springer, 2016.

\bibitem{overTheAir}
T.~J. O’Shea, T.~Roy, and T.~C. Clancy, ``Over-the-air deep learning based
  radio signal classification,'' {\em IEEE Journal of Selected Topics in Signal
  Processing}, vol.~12, no.~1, pp.~168--179, 2018.

\bibitem{branchynet}
S.~Teerapittayanon, B.~McDanel, and H.-T. Kung, ``Branchynet: Fast inference
  via early exiting from deep neural networks,'' in {\em IEEE International
  Conference on Pattern Recognition (ICPR)}, pp.~2464--2469, IEEE, 2016.

\bibitem{ACMbyML}
M.~W. Aslam, Z.~Zhu, and A.~K. Nandi, ``Automatic modulation classification
  using combination of genetic programming and knn,'' {\em IEEE Transactions on
  Wireless Communications}, vol.~11, no.~8, pp.~2742--2750, 2012.

\bibitem{cyclicAMC}
L.~Xie and Q.~Wan, ``Cyclic feature-based modulation recognition using
  compressive sensing,'' {\em IEEE Wireless Communications Letters}, vol.~6,
  no.~3, pp.~402--405, 2017.

\bibitem{2DCNN}
Y.~Sun and E.~A. Ball, ``Automatic modulation classification using techniques
  from image classification,'' {\em IET Communications}, vol.~16, no.~11,
  pp.~1303--1314, 2022.

\bibitem{Xnet}
K.~Chen, J.~Zhang, S.~Chen, S.~Zhang, and H.~Zhao, ``Automatic modulation
  classification of radar signals utilizing x-net,'' {\em Digital Signal
  Processing}, vol.~123, p.~103396, 2022.

\bibitem{withPCA}
S.~Ramjee, S.~Ju, D.~Yang, X.~Liu, A.~E. Gamal, and Y.~C. Eldar, ``Fast deep
  learning for automatic modulation classification,'' {\em arXiv preprint
  arXiv:1901.05850}, 2019.

\bibitem{EEreview}
S.~Scardapane, M.~Scarpiniti, E.~Baccarelli, and A.~Uncini, ``Why should we add
  early exits to neural networks?,'' {\em Cognitive Computation}, vol.~12,
  no.~5, pp.~954--966, 2020.

\bibitem{train1}
C.-Y. Lee, S.~Xie, P.~Gallagher, Z.~Zhang, and Z.~Tu, ``Deeply-supervised
  nets,'' in {\em Artificial Intelligence and Statistics}, pp.~562--570, 2015.

\bibitem{train2}
S.~Scardapane, D.~Comminiello, M.~Scarpiniti, E.~Baccarelli, and A.~Uncini,
  ``Differentiable branching in deep networks for fast inference,'' in {\em
  IEEE International Conference on Acoustics, Speech and Signal Processing
  (ICASSP)}, pp.~4167--4171, 2020.

\bibitem{DeeBERT}
J.~Xin, R.~Tang, J.~Lee, Y.~Yu, and J.~Lin, ``Deebert: Dynamic early exiting
  for accelerating bert inference,'' {\em arXiv preprint arXiv:2004.12993},
  2020.

\bibitem{lightAMC}
Y.~Wang, J.~Yang, M.~Liu, and G.~Gui, ``Lightamc: Lightweight automatic
  modulation classification via deep learning and compressive sensing,'' {\em
  IEEE Transactions on Vehicular Technology}, vol.~69, no.~3, pp.~3491--3495,
  2020.

\bibitem{RML22}
V.~Sathyanarayanan, P.~Gerstoft, and A.~E. Gamal, ``Rml22: Realistic dataset
  generation for wireless modulation classification,'' {\em IEEE Transactions
  on Wireless Communications}, pp.~1--1, 2023.

\bibitem{DLref}
I.~Goodfellow, Y.~Bengio, and A.~Courville, {\em Deep Learning}.
\newblock MIT Press, 2016.
\newblock \url{http://www.deeplearningbook.org}.

\end{thebibliography}

\end{document}